# Blind Interference Alignment in 6G Optical Wireless Communications

Ahmad Adnan Qidan, Member, IEEE, Taisir El-Gorashi, Member, IEEE, Jaafar M. H. Elmirghani, Fellow, IEEE

*Abstract*— In recent years, the demand for high speed wireless networking has increased considerably due to the enormous number of devices connected to the Internet. In this context, optical wireless communication (OWC) has received tremendous interest in the context of next generation wireless networks where OWC offers a huge unlicensed bandwidth using optical bands. OWC systems are directional and can naturally provide multiple-input and multiple-output (MIMO) configurations serving multiple users using a high number of transmitters in the indoor environment to ensure coverage. Therefore, multiuser interference must be managed efficiently to enhance the performance of OWC networks considering different metrics. A transmission scheme referred to as blind interference alignment (BIA) is proposed for OWC systems to maximize the multiplexing gain without the need for channel state information (CSI) at the transmitters, which is difficult to achieve in MIMO scenarios. However, standard BIA avoids the need for CSI at the cost of requiring channel coherence time large enough for transmitting the whole transmission block. Moreover, the methodology of BIA results in increased noise with increase in the number of transmitters and users. Therefore, various network topologies such as network centric (NC) and user centric (UC) designs are proposed to relax the limitations of BIA where these topologies divide the receiving area into multiple clusters. The results show a significant enhancement in the performance of topological BIA compared with standard BIA.

## I. INTRODUCTION

The massive use of wireless devices on daily basis requires a new generation (6G) of wireless communications that can provide aggregate date rates of Tb/s per access point. Furthermore, these new wireless networks must introduce features such as high achievable data rates, low latency, low power consumption, security and high reliability. Therefore, new technologies are needed in 6G to achieve these objectives including high data rate transceivers, new network architectures and efficient interference management schemes. In both industrial and academic communities, optical wireless communication (OWC) is increasingly being considered as a promising technology that can support the escalating user demands where the optical spectrum provides a huge and license-free bandwidth, satisfying part of the requirements of 6G networks. In [1], visible light communication (VLC) using light emitting diodes (LEDs) is deployed for providing illumination and communication. It is shown that VLC can outperform traditional radio frequency (RF) wireless networks in terms of achievable data rates and aggregate capacity. It is worth mentioning that there are currently significant standardization efforts for OWC in the 802.11bb standard and its integration in the Internet protocol (IP) network, usually referred to as Light-Fidelity (LiFi) [2]. However, using LEDs for data transmission limits the performance of OWC systems where these sources have a limited modulation speed. Consequently, infrared lasers, for example Vertical Cavity Surface Emitting Laser (VCSEL), can be used as alternative sources for data communication in indoor environments, since VCSELs have a high modulation speed, and can support Tb/s aggregate data rates under eye safety constraints [3], [4].

Typically, an indoor optical wireless environment contains a large number of optical access points (APs), forming a very dense optical wireless network that provides uniform data coverage. Given this point, OWC networks naturally form multiple-input and multiple-output (MIMO) systems serving multiple users. Therefore, interference management must be addressed, where all users are served simultaneously achieving a high sum rate. Orthogonal transmission schemes such as time division multiple access (TDMA), frequency division multiple access (FDMA) [5], can be used in OWC systems to avoid complexity while users are served in exclusive time or frequency slots. However, these schemes achieve low spectral efficiency. Additionally, transmit precoding (TPC) schemes such as minimizing the mean square error (MMSE) [6], zero forcing (ZF) or interference alignment (IA) [7] derived originally for RF wireless networks can potentially be applied for OWC systems in order to maximize the overall sum rate of users. However, the implementation of TPC schemes is not straightforward due to the fact that these schemes perform well when accurate channel state information (CSI) at transmitters and coordination among optical APs are fully satisfied. Besides, the characteristics of the optical transmitted signal limit the performance of TPC schemes where optical communication typically uses intensity Modulation and Direct Detection (IM/DD), which means that the transmitted signal must be non-negative, i.e. unipolar, real valued signal. On the other hand, non-orthogonal multiple access (NOMA) has been studied for OWC systems [8], in which each user with a weak channel gain is grouped with a user who experiences a strong channel gain so that their information can be superimposed in the power domain, achieving a high user rate each. However, NOMA is subject to the need for CSI, fairness and the accuracy of grouping algorithms.

In this paper, a new transmission scheme referred to as blind interference alignment (BIA) is investigated in OWC systems. In [9], BIA was proposed first for RF wireless networks in order to maximize the multiplexing gain, i.e., degrees of freedom (DoF), without the need for CSI at transmitters. Basically, BIA aligns multi-user interference by exploiting the channel correlation among users, and therefore, each user must be equipped with a reconfigurable antenna to satisfy the requirements of BIA. It was shown that BIA achieves higher DoF compared with TPC schemes in multiple-input single-output broadcast channel (MISO BC) RF wireless scenarios. In [10], a reconfigurable detector is proposed for OWC networks following the concepts of the reconfigurable antenna in order to apply BIA in OWC, while relaxing the limitations that emerge due to the nature of the optical signal including the non-negativity of the transmitted signal where the precoding matrix of BIA contains positive values only. It is worth pointing out that the performance of BIA suffers degradation in high density networks due to the requirement of long channel coherence time and the methodology of BIA in canceling the interference, more details are provided in Section III. Therefore, various network topologies such as network centric (NC) and user centric (UC) are considered to relax the BIA limitations, achieving high performance [11], [12]. In these network topologies, the receiving plane is divided into several clusters, each with unique APs and users, in order to align the interference within the area of a cluster using BIA regardless of other adjacent clusters. Simulation results of a BIA use case show that topological BIA achieves higher user rates compared to the traditional BIA scheme. Moreover, the consideration of NC and UC designs relaxes the limitation of BIA in terms of the channel coherence time required.

## II. Optical Wireless communication

In an optical wireless network, multiple optical transmitters are usually deployed on the ceiling to provide uniform coverage in an indoor environment where each optical transmitter illuminates a confined area. In this sense, OWC systems act naturally as MIMO systems, where multiple users are served at high data rates. In general, the optical channel consists of line of sight (LoS) and diffuse components, which result from the direct link and the reflections from the walls and ceiling of the room, respectively. It is worth mentioning that in the literature of OWC systems, most of the works ignore the diffuse component of the optical channel for the sake of easy channel modeling, claiming that the large portion of the received power is due to the LoS link. However, LoS components can be easily blocked, and therefore, users might experience a low SNR in environments with high blockage probabilities. In this context, the diffuse components can be leveraged to provide seamless coverage in such indoor environments, increasing the received power from most of the available transmitters. Recently, the use of intelligent reflecting surfaces (IRSs) [13] has been investigated in RF networks to ensure the connectivity of users located at a large distance from base stations. It is interesting to consider the implementation of IRSs in indoor OWC systems in order to control the radiated signals, focusing more power towards users blocked from receiving LoS components. This can also help in providing seamless transition for mobile users moving at higher speeds.

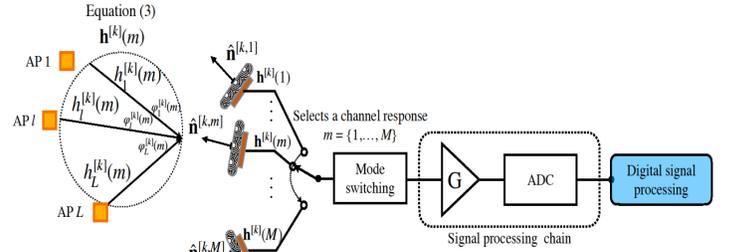

Fig. 1. Reconfigurable IR detector composed of multiple photodiodes.

### A. Vertical Cavity Surface Emitting Lasers (VCSELs)

In general, there are many types of light sources that are considered for data transmission such as regular light emitting diodes (LEDs), red, green and blue LEDs (RGB-LED) and RGB laser diodes. In particular, LEDs are commonly used in optical wireless networks for both functions, illumination and data transmission simultaneously, due to their low cost and high eye and skin safety compared with laser optical transmitters. However, LEDs are limited by their low modulation speed, which might not meet the requirements of optical wireless networks to be considered for the next generation of wireless networks. Infrared lasers such as VCSELs were considered for data transmission under eye safety constraints in [4]. It is shown that a Tb/s communication speed can be achieved. Additionally, VCSELs have several features such as low manufacturing cost, low power consumption, small size and long life time. Therefore, they can potentially be considered in the next generation of optical wireless networks to fulfill the high demands of users. VCSELs illuminate a small area, and thus a large number of VCSELs is required to ensure coverage. In this sense, an array of VCSELs must be designed to provide an extended coverage area. However, many challenges emerge in designing these arrays including the distance between two neighboring VCSELs, the number of VCSELs in an array and the transmitted power of a VCSEL, where all these issues are considered vital in determining the speed of the communication link. In this work, we consider multiple VCSELs deployed on the ceiling, serving multiple users distributed on the receiving plane. Moreover, all the VCSELs are connected to a central unit (CU) that provides time synchronization. It is worth mentioning that BIA is considered to avoid the need for CSI at transmitters, and therefore, the CU knows only the distribution of the users.

### B. Reconfigurable IR Detector

In optical wireless networks, one possible architecture attempts to ensure that each user has a wide field of view (wFoV) in order to have the possibility of being served by all the available optical wireless transmitters in the network. In addition, the characteristics of the optical channel are different than the RF channel in terms of frequency and wavelength, and therefore, the channel responses of users are highly correlated due to the lack of small scale effects. In this sense, a unique optical detector with the ability to provide wFoV is needed. Moreover, the correlation among the channel responses must be minimized to achieve high performance. Therefore, a reconfigurable detector is proposed in order to provide wFoV and linearly independent channel responses [10], [11], which are required for implementing BIA. Briefly, this detector is composed of multiple photodiodes connected to a single signal processing chain as shown in Fig. 1. Each photodiode is pointed to a distinct direction corresponding to a channel response, $\mathbf{h}^{[k]}(m)$, which is composed of channel responses $h_l^{[k]}(m)$, for user $k$ when access point $AP_l$ is used with reception mode $m$, ie photodetector $m$ is selected in user $k$ as shown in Fig. 1. Each photodetector is pointed in a different direction, thus providing a different reception mode. Also note that the signal power collected by each photodetector is affected by the angle of incidence, $\varphi^{[k]}(m)$, with respect to the photodetector normal and each photodetector has an orientation $\hat{\mathbf{n}}^{[k,m]}$ which specifies its azimuth and elevation angles [12]. Notice that, the classical angle diversity receiver, which is also composed of several photodiodes, provides almost the same channel response at each photodiode, i.e, subject to high correlation among them. The reconfigurable detector allows a user to vary its channel state over the transmission block of BIA.

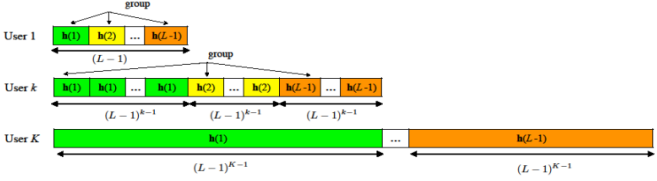

Fig. 2. The construction of Block 1. Each colour corresponds to a preset mode.

## III. IV. BLIND INTERFERENCE ALIGNMENT

In [9], BIA was proposed as a signal processing scheme for maximizing the DoF in RF wireless networks without CSI at transmitters. The methodology of BIA is based on exploiting the channel variations among users such that a set of symbol extensions must be dedicated for each user. Each symbol extension corresponds to a time slot in the time domain. Therefore, the channel state of a user varies during its time slots according to a predefined pattern of a BIA transmission block. To define the structure of this transmission block determining the channel states of each user, a reconfigurable antenna with the ability to provide a set of independent channel responses must be considered for each user. It is worth mentioning that in BIA based transmission, the physical channel must remain quasi-static over the entire transmission block of BIA. It is shown that the achievable sum-DoF for MISO BC scenarios, in which $L$ transmitters serve $K$ users, is given by $\frac{LK}{L+K-1}$. Notice that, this value is higher than the achievable DoF of orthogonal transmission schemes such as TDMA or OFDMA, and according to [9], it is considered an optimal DoF for MISO BC configurations in the absence of CSIT. In this section, the implementation of BIA in optical wireless networks is described in detail.

### A. Standard BIA

BIA aligns the desired information of each user into a full rank matrix, while interfering signals are contained in a matrix that has at least one dimension less. Interestingly, the transmission block of BIA contains multiple resource blocks allocated to each user. During a resource block of user $k$, the channel state of that user changes among $L$ distinct preset modes of its reconfigurable photodetector. While the channel states of all other users remain constant. Therefore, the transmission block of BIA guarantees the linear independence of symbols transmitted over a resource block of user $k$. In addition, the independence among multiple resource blocks allocated for a user must be ensured, giving the ability for each user to decode its information transmitted over the entire transmission block. This condition can be satisfied by transmitting the resource blocks of each user in orthogonal fashion, so that the resource blocks of each user do not overlap. In other words, the resource blocks of a user do not contain any symbol in common. Following this methodology, the independence among the desired symbols of user $k$ is ensured, and the interference received due to the transmission to other users is aligned into a matrix that has less dimensions than the useful information intended to user $k$ over its resource blocks.

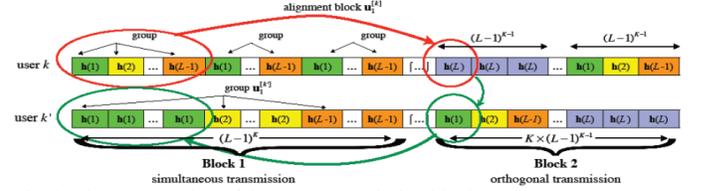

Fig. 3. The construction of the BIA transmission block.

To guarantee the methodology mentioned above, the transmission block of BIA is divided into two blocks referred to as Block 1 and Block 2. The length of each block varies based on the density of the network, where it is given by the number of APs and users. Furthermore, during Block 1, all users must be served simultaneously, while orthogonal transmission is carried out over Block 2. The design of these blocks is explained in the following, and further information can be found in [9].

#### 1) Design Block 1:
For MISO BC with $L$ optical APs and $K$ users, Block 1 consists of multiple non-overlapping groups allocated to each user. Each group corresponds to one resource block as described in Fig. 2. During a specific group of user $k$, the channel state of that user changes among $L$ preset modes given by $L$ transmitters, while the channel states of other users remain constant in order to give enough dimensions for each user to measure the interference afterwards. Consequently, the transmission block contains $(L-1)^{K-1}$ groups over Block 1 allocated for each user, satisfying the methodology of BIA. It is worth mentioning that in Block 1, user $k$ changes the preset mode selected by its reconfigurable photodetector every $(L-1)^{K-1}$ time slots over a total of $(L-1) \times (L-1)^{K-1}$ time slots, ensuring the construction of its groups in orthogonal fashion. Therefore, the independence of data streams transmitted to a user is guaranteed. Notice that, Block 1 is constructed following the same procedure for all other users. As a result, Block 1 comprises $(L-1)^K$ time slots as shown in Fig.

2. More information regarding the mathematical construction of Block 1 is provided in [9].

*2) Design Block 2*: Once Block 1 of the transmission block is built, the construction of Block 2 is straightforward. Recall that the information intended to each user must be transmitted in orthogonal fashion during Block 2 in order to measure the interference received over Block 1. Each group within Block 1 comprises $L - 1$ time slots. Therefore, an additional time slot is required to complete each resource block allocated to a user, where a resource block of a user comprises a total of $L$ time slots, over which its reconfigurable detector switches among $L$ distinct preset modes. Consequently, Block 2 has to provide $K(L-1)^{K-1}$ time slots to complete $(L-1)^{K-1}$ groups allocated for each user over Block 1. Finally, $(L-1)^{K-1}$ resource blocks are generated for each user over the entire transmission block of BIA as shown in Fig. 3.

### B. Limitations of BIA

BIA provides multiple access in optical wireless networks, where the interference is aligned without the need for CSI at transmitters. Moreover, the precoding matrix of BIA naturally satisfies the non-negativity of the transmitted signal, and its performance is less affected by the correlation among the channel responses of users compared with other transmission schemes. However, standard BIA is more suitable in networks with low density, i.e., a low number of transmitters serving multiple users. In large-size networks, BIA provides rates that decrease considerably with the number of transmitters and users. This is because of the fact that BIA is subject to limitations as follows. Firstly, the length of the BIA transmission block increases as the number of APs and users increases. Therefore, the channel coherence time must be large enough during transmission, so that the physical optical channel can remain constant while the BIA block is transmitted, which is difficult to guarantee in such networks with high density. Secondly, each user must subtract the information intended to all other users, resulting in severe noise, which increases with number of users. Therefore, in low SNR scenarios, BIA achieves low performance in terms of the achievable DoF and rate. As a consequence, the following points must be taken into consideration prior to the implementation of BIA: i) the length of the transmission block increases exponentially with the number of optical APs and users, which determines the requirements of the channel coherence time, ii) noise caused by interference subtraction increases with the number of users, iii) applying the standard BIA scheme directly to an optical wireless network results in a large static cell, in which all users are connected to the whole set of optical APs. As a result, alternative BIA schemes must be derived in order to overcome these issues. In the following, various network topologies are used in relaxing the limitations of BIA in high density optical wireless networks.

### C. Topological BIA

The implementation of the standard BIA scheme leads to forming a large static cell regardless of the distribution of users. Consequently, BIA might achieve low performance due to its limitations in terms of the length of the transmission block and the increase in noise with interference subtraction. Given this point, various network topologies can be designed to enhance the performance of BIA where the receiving plane is divided into multiple clusters. In [11], a network centric (NC) approach is designed with the aim of grouping multiple neighboring APs forming a static cluster. In this sense, BIA serves multiple users belonging to each static cluster regardless of the transmission of other clusters formed. Therefore, the NC designs enhance the performance of BIA where the length of the transmission block and the increase in noise as a result of BIA, decrease with the number of the static clusters assigned. However, the NC design has many drawbacks where users located at the edges of the NC clusters are subject to inter-cluster interference. Additionally, such static clusters are formed regardless of the distribution of users, which results in imbalance in the load among them due to the fact that some clusters may end up serving more users than others. In [11], [12], a user-centric (UC) design is implemented with the aim of forming clusters of elastic shapes. In the UC design, the clusters are formed based on the distribution of users. Therefore, the network topology is subject to updates as time progresses and users move, thus responding to changes in the network. Given this point, the inter-cluster interference can be managed as noise, enhancing the achievable sum rate of BIA within the area of each elastic cluster. An example of a network topology given by applying a UC algorithm is shown in Fig.4.

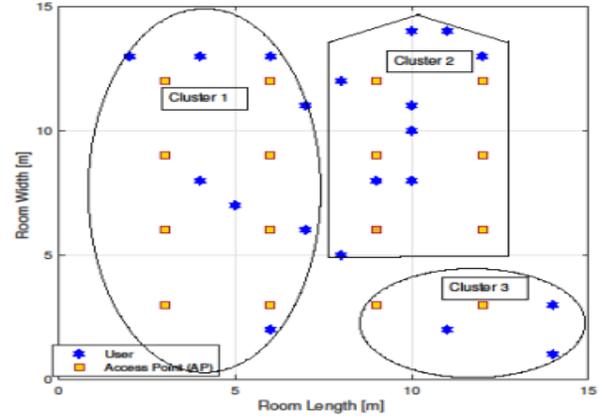

Fig. 4. Network topology given by a UC design. Three clusters are assigned.

### IV. OPTIMIZATION PROBLEMS

Formulating optimization problems is needed to enhance the performance of optical wireless networks under different metrics. Recall that BIA is limited by the high number of APs and users. Therefore, an objective function that maximizes the sum rate at a given time can be defined under several constraints that relax the limitations of BIA. Besides, optimal UC and NC approaches can be derived by formulating optimization problems that aim to group users and APs in an optimal fashion. It is worth mentioning that these optimization problems are usually NP-hard problems, where several variables are related to each other. For instance, forming an optimal UC design requires two variables to be defined that work as indicators for whether an AP and a user are assigned to each other or not taking into consideration the maximization of the achievable rate. Therefore, such an optimization problem is difficult to solve due to its high complexity where various AP-User combinations must be searched exhaustively to find the optimal network topology.

Furthermore, load balancing among multiple optical cells is also a vital solution to maximize the overall BIA-based achievable rate of the network. These optimization problems are usually defined as Mixed integer nonlinear programming (MINLP) problems, which are not tractable. Interestingly, methods such as deterministic and stochastic algorithms can be used for solving MINLP problems. In [14] an optimization solver referred to as BARON was used to solve non-convex optimization problems. However, more practical solutions are needed to avoid the high complexity of such optimization problems, while achieving sub-optimal solutions considerably close to the optimal ones as in the following.

### A. Distributed and Heuristic solutions

Distributed algorithms using full decomposition method are used for solving MINLP problems in [4], where Lagrangian functions are derived from original problem. Basically, the Lagrangian function contains multipliers, each corresponding to a constraint in the original problem. These multipliers work as bridges to synchronize the solutions of two problems solved separately on the user and AP sides, achieving sub-optimal solutions. It is worth remarking that distributed solutions require iterative algorithms that might take time while providing solutions in real-time scenarios. On the other hand, low complexity heuristic solutions solve optimization problems of high complexity with low computational time. However, heuristic algorithms provide solutions with low optimality compared with distributed solutions.

### B. Deep Learning solutions

Machine learning (ML) techniques have been considered recently as a practical solution for such optimization problems that have high complexity. In [15], ML techniques are discussed for use in solving crucial issues in wireless communication such as mobility, networking, resource management and localization where solutions are obtained without prior knowledge of the network, which is difficult to provide in real scenarios. In this sense, using ML is beneficial in finding AP-user combinations that maximize the performance of BIA in terms of the length of the transmission block and the increasing noise. For instance, the K-means ML algorithm can be used to divide elements into multiple clusters based on the closest distance. The implementation of the K-means algorithm requires first assigning a number of clusters, and then, the formation of each cluster starts by choosing a user randomly so that other users with closest distance are grouped with that user forming a unique set of users belonging to a cluster. Notice that, the K-means algorithm must be combined with an AP association technique that determines a unique set of APs for each cluster. Moreover, applying the K-means algorithm in BIA-based optical networks means that an optimal number of clusters must be searched. Therefore, traditional ML techniques might require high computational time especially in large size-networks.

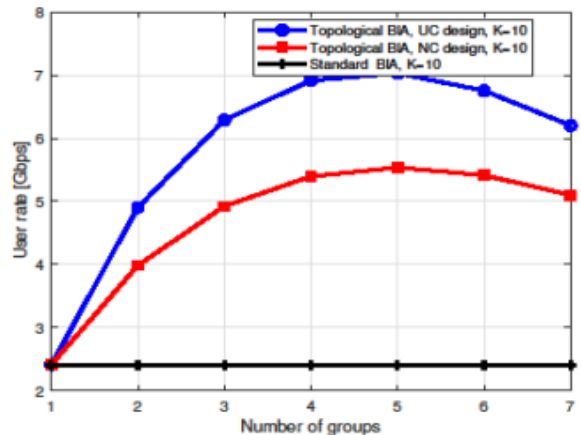

Fig. 5. The achievable user rates for the considered BIA schemes.

Finally, deep learning (DL) is a promising ML technique that uses artificial neural networks (ANNs) for learning purposes. It is worth mentioning that ANNs play a major role in making the use of DL more suitable for cellular networks compared with other ML techniques. Briefly, original optimization problems including clustering, load balancing and rate maximization problems can be solved using solvers that provide optimal or sub-optimal solution in an offline phase. The results are recorded in the form of datasets and are then used to train the model in finding optimal solutions in an online phase with low computational time. Therefore, the use of the trained model can be leveraged in the real time phase for providing instantaneous estimations for problems with different contexts. Different models of ANNs such as the multilayer perceptron (MLP) and the convolutional neural network (CNN) can be applied.

## V. BIA USE CASE

To evaluate the performance of BIA, a room with dimensions: length × width × height of 5×5×3 m, is considered, in which multiple VCSELs are deployed on the celling serving multiple users. It is worth mentioning that each VCSEL covers a confined area given the fact that the beam width is narrow [4], [12]. Moreover, users are distributed randomly on the receiving plane, and each user is equipped with a reconfigurable detector that provides a set of channel responses corresponding to the number of transmitters.

In Fig.5, the achievable user rate of BIA is depicted against the number of groups considering various network topology approaches. It is shown that topological BIA schemes achieve a user rate that increases with the number of groups from 2 to 5. However, the user rate starts decreasing considerably as the number of groups increases beyond 5. This behavior of topological BIA schemes is expected due to the trade-off between the achievable user rate and the number of groups. In other words, having a high number of groups means that the complexity of BIA is low, while severe inter cluster interference (ICI) is generated among the groups. Comparing UC and NC designs, BIA is more suitable with the UC design due to that fact that the groups are formed with elastic shapes considering the distribution of users.

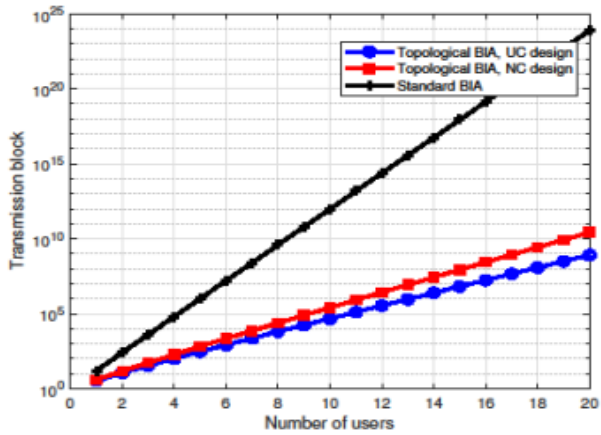

Fig. 6. The lengths of the transmission block for the considered BIA schemes.

The length of the BIA transmission block against the number of users is shown in Fig.6. It can be seen that standard BIA requires $10^{12}$ slots to serve 10 users. On the other hand, topological BIA schemes based on UC and NC designs require around $10^5$ and $10^6$ time slots, respectively, for the same number of users. It is worth mentioning that the reduction in the length of the transmission block of topological BIA schemes is due to the implementation of NC and UC designs where multiple clusters are formed with unique elements, and BIA is applied within each cluster independently. While, standard BIA aligns the interference taking into consideration the transmission to all the users in the network. The figure further shows a slight difference in the length of the transmission blocks between NC and UC approaches, which is caused by the fact that the UC design forms clusters considering the distribution of users in the network.

## VI. CONCLUSIONS

In this paper, interference management in optical wireless networks is addressed using BIA transmission schemes. In BIA, the transmission block is divided into two blocks in order to align multi-user interference blindly without the need for CSI at transmitters. The performance of BIA is limited by the length of the transmission block, which increases with the size of the network, and the inherent noise that results from interference subtraction. Therefore, topological BIA schemes are proposed with the aim of dividing the network into several clusters based on NC and UC approaches. Then, BIA is used for data transmission within each cluster regardless of the transmission of other neighboring clusters due to the fact that each cluster is composed of unique elements. The results show that the topological BIA scheme given by the UC design achieves high data rates compared with the NC design or the traditional BIA scheme. Moreover, the length of the transmission block decreases considerably when NC and UC designs are implemented.

**Ahmad Adnan-Qidan** is working as a postdoctoral research fellow with the School of Electronic and Electrical Engineering, University of Leeds, UK. His research focus is on optical wireless cellular Networks, interference management, machine learning based wireless communication and hybrid networking.

**Taisir El-Gorashi** is a Lecturer in optical networks in the School of Electronic and Electrical Engineering, University of Leeds. Her research focus is on the energy efficiency of optical networks, distributed cloud computing, network virtualization, big data and optical wireless networks.

**Prof. Jaafar Elmirghani** is Fellow of IEEE, Fellow of the IET, Fellow of the Institute of Physics and is the Director of the Institute of Communication and Power Networks and Professor of Communication Networks and Systems within the School of Electronic and Electrical Engineering, University of Leeds, UK. He has provided outstanding leadership in a number of large research projects, and was PI of the £6m EPSRC Intelligent Energy Aware Networks (INTERNET) Programme Grant, 2010-2016, and is the PI of EPSRC £6.6m Terabit Bidirectional Multi-user Optical Wireless System (TOWS) for 6G LiFi, 2019-2024. His work led to 5 IEEE standards with a focus on network virtualisation and energy efficiency, where he currently heads the work group responsible for IEEE P1925.1, IEEE P1926.1, IEEE P1927.1, IEEE P1928.1 and IEEE P1929.1. He has research interests in communication systems and networks energy efficiency and in optical wireless systems and networks.